# Magnetic Susceptibility of NiO Nanoparticles

S. D. Tiwari and K. P. Rajeev

Department of Physics, Indian Institute of Technology, Kanpur 208016, Uttar Pradesh, India

Nickel oxide nanoparticles of different sizes are prepared and characterized by x-ray diffraction and transmission electron microscopy. A.C. susceptibility measurements as a function of temperature are carried out for various particle sizes and frequencies. We find that the behavior of the system is spin glass like.

During the last few decades magnetic nanoparticles have been attracting the attention of scientists both from the view point of fundamental understanding as well as applications[1]. Magnetic nanoparticles have been a subject of interest since the days of Néel[2] and Brown[3] who developed the theory of magnetization relaxation for noninteracting single domain particles. Effects of interparticle interactions on magnetic properties of several nanoparticle systems have been reported by several authors[4]. In 1961 Néel suggested that small particles of an antiferromagnetic material should exhibit magnetic properties such as superparamagnetism and weak ferromagnetism[5]. Antiferromagnetic nanoparticles show more interesting behavior compared to ferro and ferrimagnetic nanoparticles. One such interesting behavior is that the magnetic moment of tiny antiferromagnetic particles increases with increasing temperature[6,7] which is quite unlike what is seen in ferro or ferrimagnetic particles. Real magnetic nanoparticles have disordered arrangement, distribution in size and random orientation of easy axes of magnetization, making their behavior very complex and enigmatic.

Bulk NiO has rhombohedral structure and is antiferromagnetic below 523 K whereas it has cubic structure and is paramagnetic above that temperature[8]. A bulk antiferromagnet has zero net magnetic moment in zero applied field. If the surface to volume ratio, which varies as the reciprocal of particle size, for antiferromagnetic particles becomes sufficiently large then the particles can have nonzero net magnetic moment because of uncompensated spins at the surface and the magnetic properties of a collection of these particles can be very different from that of the corresponding bulk material. According to Néel the moment due to the uncompensated spins would be parallel to the axis of the antiferromagnetic alignment. Richardson and Milligan[9] were the first to report magnetic susceptibility measurements as a function of temperature for NiO nanoparticles of different sizes. They found a peak in the susceptibility data much below the Néel temperature and the peak temperature ($T_C$) was found to decrease with decrease in particle size. They attributed the observed behavior to the modified magnetic environment of the $Ni^{+2}$ ions at the surface of the particles and the enhanced surface to volume ratio. In this paper we present and analyze the particle size and frequency dependence of a.c. susceptibility vs. temperature measurements on NiO nanoparticles.

NiO nanoparticles are prepared by a sol-gel method by reacting in aqueous solution, at room temperature, nickel nitrate and sodium hydroxide at pH = 12 as described elsewhere[9, 10, 11]. In this work we used nickel (II) nitrate hexahydrate (99.999%), and sodium hydroxide pellets (99.99%), both from Aldrich, and triple distilled water to make nickel hydroxide gel. The samples of nickel oxide nanoparticles are prepared by heating the nickel hydroxide gel at a few selected temperatures for 3 hours in flowing helium gas (99.999%). All the magnetic measurements are done with a SQUID magnetometer (Quantum Design, MPMS XL).

The XRD patterns of the nickel oxide samples taken at the relatively low speed of 0.6 degree/minute, using a Seifert diffractometer with CuKα radiation, are shown in Fig.1. The gradual broadening of the peaks as the temperature of heat treatment is decreased confirms the formation of nanoparticles. The average crystallite size is calculated by x-ray diffraction line broadening using the Scherrer formula[12] $t = 0.9 \lambda / \cos\theta_B \sqrt{(B_M^2 - B_S^2)}$ where $\lambda$ is the wavelength of the x-rays (1.542 Å), $2\theta_B$ is the Bragg angle, $B_M$ is the full width at half maximum (FWHM) of a peak in radians and $B_S$ is the FWHM of the same peak of a standard sample for which we used a bulk sample of "Specpure" grade NiO (Johnson Matthey & Co. Ltd., UK). Peaks (111), (200) and (220) are used to calculate the average crystallite size. The use of $\sqrt{(B_M^2 - B_S^2)}$ instead of $B_M$ in the Scherrer formula takes care of instrumental broadening. The crystallite sizes of NiO samples prepared by heating $Ni(OH)_2$ at 250, 300, 350 and 700 $^0$C, turn out to be 5.1, 6.2, 8.5 and >100 nm respectively and these numbers will be referred to as the average crystallite size in this paper. The XRD patterns are taken thrice for each sample to check the reproducibility and we found that the average crystallite size determined from different data sets did not differ by more than 0.1 nm. Transmission electron micrograph (TEM), the corresponding selected area diffraction (SAD) pattern and the particle size distribution of the sample prepared by heating Ni(OH)$_2$ at 250 $^0$C are shown in figures 2a, 2b and 2c respectively. From figure 2a it is clear that the particles are of arbitrary shape. The mean particle size was found to be 5.6 nm with a standard deviation of 1.3 nm. We notice that the mean particle size determined by TEM is very close to the average crystallite size determined by XRD which implies that on the average each NiO nanoparticle is a crystallite. The radius of a ring in SAD pattern is proportional to $\sqrt{(h^2 + k^2 + l^2)}$ where h,k,l are the Miller indices of the planes corresponding to the ring. Using this information we confirmed that the electron diffraction pattern shown in figure 2b is that of NiO.

In Fig.3 we report the dc-susceptibility as a function of temperature for all the samples measured in a field of 3500 G. Richardson and Milligan[9] had reported the same measurements on similarly prepared samples and we note that our data agree with their data very closely showing that the results are reproducible. The most important features we note in figure 3 are that (i) there is a peak in the susceptibility data and the peak temperature is much lower than the Néel temperature of the bulk material (ii) the susceptibility increases substantially as the particle size decreases and reaches a few nanometers (iii) the temperature of the peak decreases with decreasing particle size.

We measured the ac-susceptibility in the frequency range 1 Hz to 1 kHz in decade steps. The sample is first cooled from room temperature to 10 K in a zero magnetic field. A probing a.c. magnetic field of 1.0 G amplitude is used to measure the susceptibility as the temperature is slowly raised in short steps to 300 K.

Fig.4 shows the real ($\chi'$) and imaginary ($\chi''$) parts of the a.c. susceptibility of the 5.1 nm sample. It is clear that that the value of $\chi'$ decreases and peak temperature increases as the frequency is increased. These are characteristics of superparamagnets and spin glasses. The inset shows that below the peak temperature, $\chi''$ does not depend on frequency. Such a behavior is known to be related to interparticle interactions, which slows down the relaxation at low temperature[13]. Fig.5 shows a comparison of $\chi'$ for 5.1, 6.2, 8.5 and > 100 nm samples at 10 Hz. It is clearly seen that the peak temperature decreases with increase in particle size which is in marked contrast to what is seen at the higher field of 3500 G (Fig. 3).

In the case of a superparamagnet the blocking temperature increases with increase in particle size because the energy barrier separating the two low energy states for a superparamagnetic particle is proportional to the volume of the particle. Fig. 5 shows that the peak temperature in $\chi'$ vs. T curve decreases with increase in particle size. Therefore the observed behavior of the sample at low field cannot be because of superparamagnetism. A quantitative measure of frequency shift is the relative shift in peak temperature ($\Delta T_f/T_f$) per decade of frequency. For 5.1 nm sample this quantity turns out to be 0.02. For well studied spin glasses it lies between 0.0045 and 0.06 whereas for a known superparamagnet a-$(Ho_2O_3)(B_2O_3)$ it has a value of 0.28[14]. The size of this quantity suggests that the peak in $\chi'$ vs. T might be because of spin glass like freezing and not because of superparamagnetic blocking of particle magnetic moments.

If we assume that there is a phase transition in the system, the spin freezing can be analyzed in terms of a critical slowing down above $T_C$, the critical temperature. The results of dynamical scaling relates the critical relaxation time $\tau$ to the correlation length $\xi$ as $\tau \sim \xi^z$, where z is a constant. Since $\xi$ diverges with temperature as $\xi \sim [T/(T-T_c)]^\nu$, we have the power-law divergence $\upsilon = \upsilon_0 \,[(T-T_C)/T]^{z\nu}$,[14] where z$\nu$ is called the dynamical exponent. By fitting the data to this power law we get $\upsilon_0 \approx 10^{12}$ Hz, z$\nu \approx 11$ and $T_C \approx 148$ K which should correspond to the d.c. (equilibrium) value of freezing temperature $T_f$ ($\upsilon \rightarrow 0$). Values of these fitting parameters are not unphysical. For spin glasses the value of z$\nu$ is known to lie between 4 and 12[14]. From this we conclude in the case of NiO nanoparticles we are most likely seeing a spin glass like freezing. If this indeed is the case, now the question is, what can cause the freezing? We consider some of the possible mechanisms below:

CASE 1 - *Interaction among particles giving rise to spin glass like behavior*: There are reports in the literature where dipolar interaction between particles has been proposed as the reason for the freezing of particle magnetic moments[15]. Now we would like to examine whether there is such a possibility in the case of NiO. We estimate that for the 5.1 nm particles there would be an average uncompensated moment of the order of 100 $\mu_B$[16]. The maximum dipolar interaction energy between two such particles sitting close to each other would be $\sim 10^{-17}$ erg which corresponds to about 0.1 K on temperature scale. This means that if dipolar interaction were causing the spin freezing it would occur at 0.1 K which is much lower than the observed freezing temperature of about 161 K. Thus we rule out the possibility that dipolar interaction among particles is causing the peaks in the $\chi'$ vs. T curves.

Now let us consider whether the exchange interaction between the surface spins of neighboring particles can play any role. If the particles are sufficiently close to each other then exchange interaction among surface spins of randomly oriented particles can give rise to disorder and frustration and this may lead to a spin glass like phase. For smaller particles the total area of contact among particles can be expected to be large compared to that for bigger particles, giving rise to more exchange interaction energy and that could possibly account for the increasing freezing temperature as particle size decreases.

CASE 2 - *Interactions within a particle giving rise to spin glass like behavior:* Kodama et al.[17] have proposed a model for ferrimagnetic nickel ferrite nanoparticles where the spins at the surface of a particle are disordered, leading to frustration and spin glass like freezing. A similar model has also been proposed for ferrihydrite nanoparticles[6]. We feel that such a model may be applicable in the case of NiO nanoparticles also. We will first have to see whether we can propose any mechanism for surface spin disorder in our case. The exchange interaction between two neighboring $Ni^{2+}$ ions is mediated by an oxygen ion (superexchange), and if an oxygen ion is missing from the surface, the exchange bond would be broken and the interaction energy would be reduced. Coordination number at the surface for $Ni^{2+}$ ions will be less than in the bulk and this can result in a distribution of exchange interaction energies for the surface spins. Also the superexchange is sensitive to bond angles and bond lengths, which are likely to be different at the surface compared to the bulk. The reason mentioned above may be sufficient to give rise to surface spin disorder and frustration leading to a spin glass like phase. The fraction of atoms lying on the surface of the particles increases as the particle size decreases. This may lead to increased surface spin disorder as the particle size decreases and could possibly account for the increasing freezing temperature as the particle size decreases.

By increasing the separation between the particles by some means it would be possible to test whether the observed behavior is because of interparticle (case 1) or intraparticle (case 2) interactions.

Now we should consider the behavior of the system at 3500 G which is at variance with the 1 G data as made clear in figures 3 and 5. We have shown that the 1 G behavior is consistent with what is expected from a spin glass like system. At 3500 G it is likely that the spin glass phase is broken because of the high field and this may be the reason we find a different behavior. The observed 3500 G behavior cannot be attributed to superparamagnetism because the variation of the peak temperature is not in accordance with the variation of the particle volume, as would be expected in the case of a superparamagnet. Can we ascribe the observations at 3500 G to a decrease in Néel temperature with decreasing particle size as in the case of $FeF_2$ nanoparticles[18]? This can also be ruled out because the neutron diffraction data[9] indicate that the NiO nanoparticles are antiferromagnetic both above and below the peak temperature of the susceptibility.

In summary, average crystallite size calculated from Scherrer formula and mean particle size determined from TEM are very close to each other suggesting that each NiO particle is a crystallite. At low field the peak temperature in the susceptibility vs. temperature curve increases as the particle size decreases. At high field the peak temperature increases as the particle size increases. The behavior of sample in low field is found to be spin glass like.

We thank Prof. A. K. Majumdar and Prof. A. Singh for useful discussions.

**Figure Captions**

Fig. 1. Room temperature XRD patterns of NiO nanoparticles prepared by heating Ni(OH)$_2$ at different temperatures shown for 3 hours in stream of pure helium gas. The crystallite size shown is the average value determined from the (111), (200) and (220) peaks using the modified Scherrer formula.

Fig. 2. (a)Transmission electron micrograph of NiO sample prepared by heating Ni(OH)$_2$ at 250 $^0$C. (b) shows Bragg diffraction from the NiO particles. Counting from the centre I$^{st}$, II$^{nd}$, III$^{rd}$….. rings correspond to (111), (200), (220)….. planes respectively. The IV$^{th}$ ring actually consists of two rings very close to each other corresponding to planes (311) and (222). (c) shows the distribution of particle sizes. This histogram is based on size measurement of a total of 170 particles. The histogram peaks at 5 nm.

Fig. 3. d.c. susceptibility of NiO nanoparticles in a field of 3500 G. The dotted line drawn to pass through the peaks shows that the peak temperature increases with increasing particle size.

Fig. 4. Temperature and frequency dependence of real part of a.c. susceptibility of 5.1 nm NiO nanoparticles. The inset shows the temperature and frequency dependence of the imaginary part of the a.c. susceptibility. The probing field has an amplitude of 1 G.

**Fig. 5.** Temperature variation of the real part of a.c. susceptibility of NiO nanoparticles in an a.c. field of amplitude 1 G and frequency 10 Hz. The dotted line drawn to pass through the peaks shows that the peak temperature increases with decreasing particle size.

**Figures**

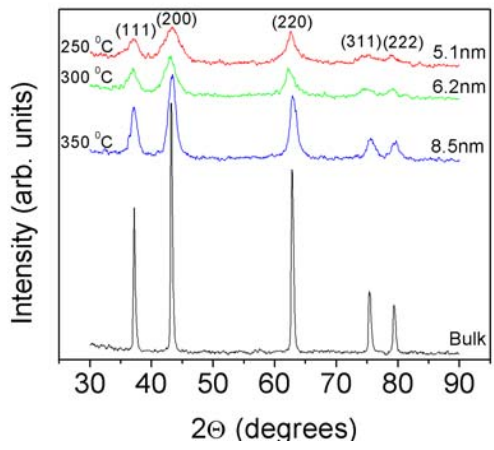

Fig. 1, S. D. Tiwari

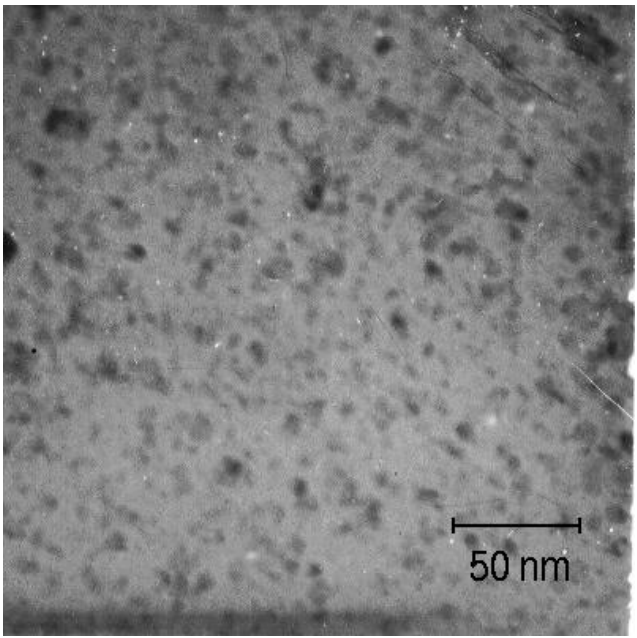

Fig. 2a, S. D. Tiwari

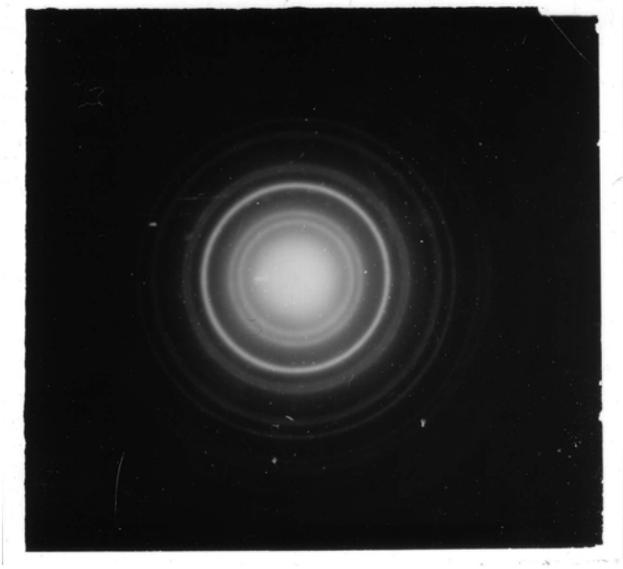

Fig. 2b, S. D. Tiwari

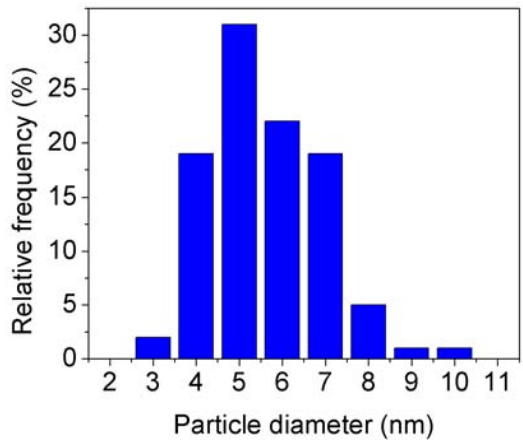

Fig. 2c, S. D. Tiwari

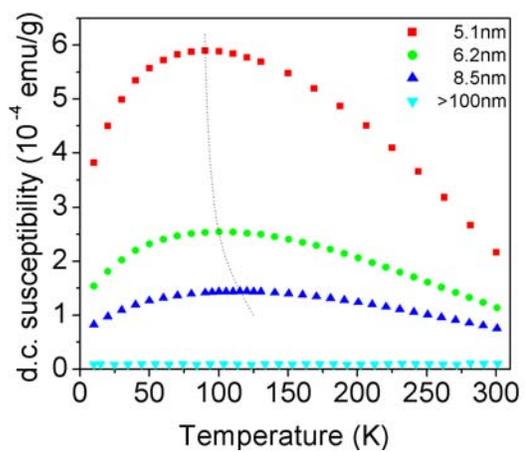

Fig. 3, S. D. Tiwari

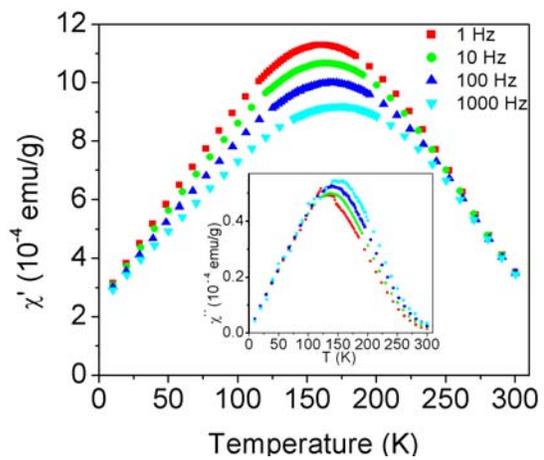

Fig. 4, S. D. Tiwari

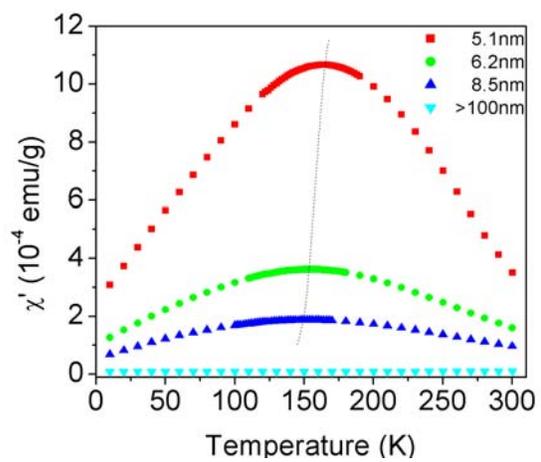

Fig. 5, S. D. Tiwari